\batchmode
\makeatletter
\def\input@path{{/Users/ananyabalakrishna/Desktop/}}
\makeatother
\documentclass[10pt,twocolumn]{article}
\usepackage[latin9]{inputenc}
\usepackage[letterpaper]{geometry}
\usepackage{color}
\usepackage{array}
\usepackage{amsmath}
\usepackage{amssymb}
\usepackage{graphicx}
\usepackage{setspace}
\usepackage{subscript}
\usepackage[unicode=true,pdfusetitle,
 bookmarks=true,bookmarksnumbered=false,bookmarksopen=false,
 breaklinks=false,pdfborder={0 0 1},backref=false,colorlinks=false]
 {hyperref}

\makeatletter

\DeclareFontEncoding{LGR}{}{}
\DeclareRobustCommand{\greektext}{%
  \fontencoding{LGR}\selectfont\def\encodingdefault{LGR}}
\DeclareRobustCommand{\textgreek}[1]{\leavevmode{\greektext #1}}
\ProvideTextCommand{\~}{LGR}[1]{\char126#1}

\providecommand{\tabularnewline}{\\}

\usepackage{ragged2e}
\usepackage{booktabs}
\usepackage{authblk}
\usepackage{setspace}
\usepackage{titlesec}
\titlespacing*{\section}{0pt}{1.1\baselineskip}{\baselineskip}

\usepackage[parfill]{parskip}
\usepackage{graphicx}
\usepackage{cite}
\usepackage{notoccite}

\title{Nanoscale domain patterns and a concept for an energy harvester}
\author{Ananya Renuka Balakrishna, John E. Huber}
\affil{Department of Engineering Science, University of Oxford, Oxford, OX1 3PJ, United Kingdom}

\date{}                    

\makeatother

\begin{document}
\twocolumn[   
\begin{@twocolumnfalse}  
\begin{center}
\textbf{\LARGE{}Li-diffusion accelerates grain growth in intercalation
electrodes: a phase-field study}
\par\end{center}{\LARGE \par}

\bigskip{}

\begin{center}
Ananya Renuka Balakrishna{*}, Yet-Ming Chiang and W. Craig Carter
\par\end{center}

\medskip{}

\begin{center}
{\footnotesize{}Department of Materials Science and Engineering, Massachusetts
Institute of Technology, Cambridge, MA02139, USA}
\par\end{center}{\footnotesize \par}

\smallskip{}

\begin{center}
{\scriptsize{}$^{*}$Email: ananyarb@mit.edu}
\par\end{center}{\scriptsize \par}

\bigskip{}

\begin{abstract}
\textcolor{black}{Grain boundary migration is driven by the boundary's
curvature and external loads such as temperature and stress. In intercalation
electrodes an additional driving force results from Li-diffusion.
That is, Li-intercalation induces volume expansion of the host-electrode,
which is stored as elastic energy in the system. This stored energy
is hypothesized as an additional driving force for grain boundaries
and edge dislocations. Here, we apply the 2D Cahn-Hilliard \textendash{}
phase-field-crystal (CH-PFC) model to investigate the coupled interactions
between highly mobile Li-ions and host-electrode lattice structure,
during an electrochemical cycle. We use a polycrystalline FePO$_{4}$
/ LiFePO$_{4}$ electrode particle as a model system. We compute grain
growth in the FePO$_{4}$ electrode in two parallel studies: In the
first study, we electrochemically cycle the electrode and calculate
Li-diffusion assisted grain growth. In the second study, we do not
cycle the electrode and calculate the curvature-driven grain growth.
External loads, such as temperature and stress, did not differ across
studies. We find the mean grain-size increases by $\sim11\%$ in the
electrochemically cycled electrode particle. By contrast, in the absence
of electrochemical cycling, we find the mean grain-size increases
by $\sim2\%$ in the electrode particle. These CH-PFC computations
suggest that Li-intercalation accelerates grain-boundary migration
in the host-electrode particle. The CH-PFC simulations provide atomistic
insights on diffusion-induced grain-boundary migration, edge dislocation
movement and triple-junction drag-effect in the host-electrode microstructure.}

\medskip{}

\noindent \textit{\footnotesize{}Key words}: grain growth, phase-field
crystal model, intercalation electrodes, batteries
\end{abstract}
\bigskip{}

\bigskip{}

\end{@twocolumnfalse} ]
\begin{singlespace}

\section*{{\small{}Introduction}}
\end{singlespace}

\begin{singlespace}
\noindent \textcolor{black}{\small{}An ion storage electrode used
in rechargeable batteries typically consists of polycrystalline (secondary)
particles. The electrode's volume can substantially expand (e.g.,
up to $\sim10\%)$ as the working ion concentration is varied \cite{Kai_YMC}.
This expansion is often anisotropic \cite{Yuan_et_al_2015}, and sometimes
leads to mechanical fracture of the electrode \cite{Woodford}. Mechanical
fracture impacts a battery's electrochemical properties, such as accessible
capacity, rate capability, and lifetime \cite{Ping-Chun-Bohua}. An
electrode's physical properties, such as fracture toughness and electrochemical
potential, correlate with its grain size in polycrystalline particles
\cite{Swallow},\cite{Bates_Dudney_2000}. Thermal loads, as well
as dopants, are known to affect grain size \cite{Yttria_GB_mobility,Li_Fu_ZrO2}.
At elevated temperatures, in-situ electric currents also cause grain
growth in electrochemical systems \cite{Yttria_GB_mobility}\cite{YSZ_GB_mobility}.
To date, we are not aware of any work that explores how cyclic intercalation
affects grain growth in a battery electrode at ambient temperatures.
Here we examine whether \textendash{} and, if so, to what extent \textendash{}
electrochemical cycling enhances grain growth, using FePO$_{4}$ /
LiFePO$_{4}$ as a model system. }{\small \par}
\end{singlespace}

\textcolor{black}{\small{}During an electrochemical cycle, Li-ions
intercalate into a host electrode, such as FePO$_{4}$ / LiFePO$_{4}$
\cite{Wang_Wang_LFP_2014}. This insertion and extraction of Li-ions
(diffusing species) produces lattice strains in the host material,
and transforms lattice structure at critical values of Li-compositions.
In a polycrystalline electrode, these lattice transformations generate
an inhomogeneous strain field \cite{Yuan_et_al_2015}. In phase separating
materials, such as the FePO$_{4},$ an additional strain arises between
the lithiated and delithiated phases. Handwerker et al. \cite{Handwerker_Cahn,WCC_Handwerker}
propose that these internal strain fields drive grain boundaries in
the host electrode. In the present work, we apply a Cahn-Hilliard
\textendash{} phase-field-crystal model to demonstrate this diffusion-induced
grain-boundary migration in an FePO$_{4}$ host-electrode. }{\small \par}

\textcolor{black}{\small{}Researchers have developed several theoretical
models to investigate electrode microstructures in lithium batteries
\cite{12 Carter_Tang,13b  Tang_Carter-1,13 Meethong_Carter_Chiang_2007,14 Bai_Bazant_2011,15 Cogswell_Bazant_2012,16 Srinivasan_Newman,17 Ceder_phase_field_models,18 Moriwake_2013,19 Ceder_atomic,20 Lu_group_lattice_distortions,21 Islam_surface_morphologies,22 Diffusion-mechanism-MD,23 ionic-diffusion-MD}.
Based on length scales, these theroetical models can be broadly clustered
into continuum and atomistic methods. The continuum methods, such
as the phase field models, describe the time evolution of electrode
microstructures. The phase field models have been applied to investigate
the role of electrode-particle size on miscibility gap \cite{13 Meethong_Carter_Chiang_2007},
phase separation \cite{14 Bai_Bazant_2011} and stress generation
\cite{Huttin_Kamlah_2012}. A crystallographic phase-field model builds
on the classic phase-field approach by introducing grain orientation
as an additional order parameter \cite{KWC_model}. The continuum
models are developed to consider electrochemistry \cite{17 Ceder_phase_field_models},
surface morphologies \cite{21 Islam_surface_morphologies}, and coherency
strains \cite{13b  Tang_Carter-1,15 Cogswell_Bazant_2012}. These
methods, however, do not model individual lattice distortions. The
microstructural defects such as grain boundaries are empirical parameters
in these continuum methods.}{\small \par}

\textcolor{black}{\small{}Atomistic models such as the first principle
methods \cite{19 Ceder_atomic} and molecular dynamics models \cite{22 Diffusion-mechanism-MD,23 ionic-diffusion-MD},
provide insights on electrochemical stability of ionic compounds \cite{19 Ceder_atomic},
lattice distortions \cite{18 Moriwake_2013,20 Lu_group_lattice_distortions}
and ionic diffusion in electrodes \cite{22 Diffusion-mechanism-MD}.
However, these approaches are limited in length and time scales. For
example, the molecular dynamics method can typically model few hundreds
of atoms. Consequently, microstructural features such as grain boundaries
are computationally expensive to simulate using these atomistic approaches.
Furthermore, the time scales involved in molecular dynamics computations
are over thermal-vibrations. These time scales limits us to model
continuum Li-diffusion. Accelerated molecular dynamics methods use
the transition state theory to overcome the time scale limitation
\cite{AFVoter_2002}, and have been applied to simulate crystal growth
\cite{AFVOter_Crystal_growth_2001}.}{\small \par}

\textcolor{black}{\small{}Alternatively in our recent work \cite{Balakrishna_Carter_2017},
we combined a Cahn-Hilliard model with a phase-field-crystal model
(CH-PFC). This model describes phase transitions and interstitial
diffusion in a 2D theoretical framework. In this multiscale approach,
the PFC equation models the coarse-grained lattice structure of the
host-electrode material (atomistic scale). The CH equation models
Li-intercalation in the electrode (continuum scale). The coupled CH-PFC
model captures the host-electrode lattice transformations induced
from Li-diffusion. This modeling approach enables us to investigate
the time evolution of lattice arrangements in electrode particles.
For example, the model simulates dislocation movement and grain-boundary
migration in electrodes during an electrochemical cycle.}{\small \par}

\textcolor{black}{\small{}In this paper, we apply the 2D CH-PFC methods
to model Li-intercalation in an FePO$_{4}$ electrode particle. First,
we calibrate the CH-PFC model to describe the lattice geometries of
FePO$_{4}$ and LiFePO$_{4}$. Using the model, we simulate a polycrystalline
FePO$_{4}$ electrode particle. We next compute two parallel simulations:
In the first, we electrochemically cycle the electrode particle where
Li-ions intercalate into FePO$_{4}$ lattices. In the second study,
we do not cycle the FePO$_{4}$ electrode particle. We compare the
grain boundary migration in the two studies. We hypothesize that Li-intercalation
accelerates grain-boundary migration in the cycled electrode. Throughout,
we compare the CH-PFC results with the existing literature on microstructures
in electrode materials.}{\small \par}
\begin{singlespace}

\section*{{\small{}Cahn-Hilliard \textendash{} phase-field-crystal model }}
\end{singlespace}

\textcolor{black}{\small{}In this section, we briefly explain the
CH-PFC model applied to FePO$_{4}$ electrodes. The details of the
model are given in the appendix and in Ref. \cite{Balakrishna_Carter_2017}.
For our purposes, we note that the CH-PFC model has two order parameters:
First, the peak-density field $\psi(\vec{x})$ that represents coarse-grained
lattices of the host electrode \cite{Elder_Grant_2004}. The lattice
points of the host electrode are constrained in space and do not hop
or migrate. These lattices, however, undergo displacive transformations
to represent lattice strains. Second, the composition field $c(\vec{x})$
that represents Li-intercalation (mobile species) through the host
electrode. }{\small \par}

\textcolor{black}{\small{}During electrochemical cycling, the FePO$_{4}$
electrode typically phase-separates into lithium-rich (LiFePO$_{4}$)
and lithium-deficient (FePO$_{4}$) phases \cite{Wang_Wang_LFP_2014}.
This FePO$_{4}$ / LiFePO$_{4}$ phase transition is first order (with
an abrupt change of bulk lattice constants). In experiments, however,
an interphase region Li$_{X}$FePO$_{4}$ with $0<X<1$ has been observed
to form \cite{staging_FPLFP,Nakamura_Ikuhara_2014}. This interphase
region reduces the lattice misfit at the FePO$_{4}$ / LiFePO$_{4}$
contact \cite{staging_FPLFP}. In the current work, we approximate
the interphase region to correspond to a diffuse phase boundary. We
model the lattices in the interphase region to have averaged lattice
parameters intermediate to FePO$_{4}$ and LiFePO$_{4}$ lattices.
Both the interstitial Li-composition $c(\vec{x})$ and the host-electrode
lattice structure $\psi(\vec{x})$ continuously change across the
diffuse interface. The CH-PFC model describes the normalized total
free energy $\mathcal{F}$ of a two-phase system as:}{\small \par}

\noindent \textcolor{black}{\small{}
\begin{align}
\mathcal{F} & =\int\{g(c)+|\nabla c|^{2}+\gamma[f(r,\psi)+\frac{\psi}{2}G(\nabla_{c}^{2})\psi]\}d\vec{x.}\label{eq:1}
\end{align}
}{\small \par}

\textcolor{black}{\small{}In Eq. \ref{eq:1}, $c(\vec{x})$ is the
fraction of interstitial sites in FePO$_{4}$ electrode occupied by
Li per unit volume. The polynomial $g(c)$ represents a regular solution
model \cite{Hillert_Staffansson,12 Carter_Tang} that describes a
double well potential. The coefficients in $g(c)$ are normalized
such that the double wells are at equal heights with minima at $c(\vec{x})=0$
and $c(\vec{x})=1$. The two minimas correspond to the delithiated
(FePO$_{4}$) and lithiated (LiFePO$_{4}$) phases, respectively.}\footnote{\textcolor{black}{\footnotesize{}The $X$ in Li$_{X}$FePO$_{4}$
and $c(\vec{x})$ are related by normalization coefficients that rescale
the double well potential described by $g(c)$. }}\textcolor{black}{\small{} The gradient term $|\nabla c|^{2}$ is
the penalty for the changing Li-composition field across the FePO$_{4}$
/ LiFePO$_{4}$ diffuse phase boundary. The peak density field $\psi(\vec{x})$
is a time-averaged density field \cite{Elder_Grant_2004}, which in
its periodic state describes the coarse-grained lattice geometry of
the electrode material. The term $\frac{\psi}{2}G(\nabla_{c}^{2})\psi$
is the energy penalty resulting from the changing host-lattice structure
across the diffuse phase boundary. The difference in lattice geometries
between the FePO$_{4}$ / LiFePO$_{4}$ phases gives rise to a coherency
strain across the phase boundary. The operator $G(\nabla_{c}^{2})$
introduces the coupling between Li-composition $c(\vec{x})$ and host-electrode
lattice symmetry $\psi(\vec{x})$. The constant $\gamma$ relates
the free energy normalizations of the Cahn-Hilliard and phase field
crystal models, and is discussed in the appendix.The value of $r(\vec{x})$
in $f(r,\psi)$ affects the type of solutions for the peak density
field $\psi(\vec{x})$. In the PFC literature, researchers have used
$r(\vec{x})$ as a proxy for temperature to model glassy to non-glassy
(crystalline) transition during crystallization \cite{Emmerich_review,Ofori-Opoku_Provatas_2013}.
In this paper, we use $r(\vec{x})$ as a proxy for the amorphous Li-reservoir
in our model. For example, with $r(\vec{x})=-0.2$, $\psi(\vec{x})$
is a constant at equilibrium and Eq. \ref{eq:1} models an amorphous
state. While for $r(\vec{x})=+0.2$, $\psi(\vec{x})$ has a periodic
wave-form solution at equilibrium that describes a crystalline state
for the electrode material. The value of $r(\vec{x})$ is locally
defined in the coordinate space to model the electrode-reservoir system.
There is a surface energy contribution resulting from the difference
between the crystalline electrode and amorphous reservoir regions.
In this paper, the CH-PFC model does not account for the surface energy
term. This surface energy contribution is important to investigate
the wetting effects on phase transitions \cite{Cahn_wetting_theory,Tang_Carter_etal_2010}
and is a subject of future study.}{\small \par}

\textcolor{black}{\small{}The Li-composition field is coupled to the
electrode lattice geometry via the coordinate transformation coefficients
of the Laplace operator $\nabla_{c}^{2}$. Both the FePO$_{4}$ (FP)
and LiFePO$_{4}$ (LFP) lattices have an orthorhombic symmetry. There
are two independent variables, $\alpha$ and $\beta$, which control
FP / LFP lattice transformation. The variables, $\alpha$ and $\beta$,
interpolate the lattice parameters as a function of the composition
field, $\alpha(c)=\alpha_{\mathrm{FP}}+(\alpha_{\mathrm{LFP}}-\alpha_{\mathrm{FP}})c$
and $\beta(c)=\beta_{\mathrm{FP}}+(\beta_{\mathrm{LFP}}-\beta_{\mathrm{FP}})c$.
Here, $(\alpha_{\mathrm{FP}},\beta_{\mathrm{FP}})$ and $(\alpha_{\mathrm{LFP}},\beta_{\mathrm{LFP}})$
correspond to the geometric measurements of FePO$_{4}$ and LiFePO$_{4}$
lattices from Table \ref{tab:1}. With Li-composition $c=0$ and $c=1$,
the transformation coefficients correspond to the lattice parameters
of FePO$_{4}$ and LiFePO$_{4}$ respectively. The Li-composition
across the diffuse phase boundary is a function $c(\vec{x})$ and
the lattice is transformed according to $\alpha(c(\vec{x}))$ and
$\beta(c(\vec{x}))$. The width of the diffuse phase boundary $c(\vec{x})$
is numerically calibrated such that it spans over $\sim4$ lattice
spacings described by $\psi(\vec{x})$. }{\small \par}

\noindent \textcolor{black}{\small{}\setlength{\extrarowheight}{8pt}}{\small \par}

\noindent \textcolor{black}{\small{}}
\begin{table}
\begin{centering}
\textcolor{black}{\small{}}%
\begin{tabular}{ccccc}
\hline 
 &
\textcolor{black}{\small{}$\mathrm{a}_{2}(\textrm{Å})$} &
\textcolor{black}{\small{}$\mathrm{a}_{3}(\textrm{Å})$} &
\textcolor{black}{\small{}$\alpha=\frac{\mathrm{a_{2}}}{\mathrm{a}_{0}}$} &
\textcolor{black}{\small{}$\beta=\frac{\mathrm{a_{3}}}{\mathrm{a}_{0}}$}\tabularnewline
\hline 
\textcolor{black}{\small{}FePO$_{4}$ (FP)} &
\textcolor{black}{\small{}$9.821$} &
\textcolor{black}{\small{}$\mathrm{a_{0}}=4.788$} &
\textcolor{black}{\small{}$2.0512$} &
\textcolor{black}{\small{}$1$}\tabularnewline
\textcolor{black}{\small{}LiFePO$_{4}$ (LFP)} &
\textcolor{black}{\small{}$10.334$} &
\textcolor{black}{\small{}$4.693$} &
\textcolor{black}{\small{}$2.1583$} &
\textcolor{black}{\small{}$0.9802$}\tabularnewline
\hline 
\end{tabular}
\par\end{centering}{\small \par}
\textcolor{black}{\small{}\caption{{\footnotesize{}\label{tab:1}A list of lattice parameters $(\mathrm{a}_{2},\mathrm{a}_{3})$
for FePO$_{4}$ and LiFePO$_{4}$ used in the CH-PFC simulations \cite{Padhi_Goodenough_1997}.
The variable $\alpha,\beta$ are calculated using $\mathrm{a_{0}}=\mathrm{a_{3}{}_{(FePO_{4})}=4.788\mathring{A}}$
as reference. These coefficients describe rectangular geometries of
FePO$_{4}$ and LiFePO$_{4}$ lattices at $c=0$ and $c=1$ respectively.}}
}{\small \par}
\end{table}
{\small \par}

\noindent \textcolor{black}{\small{}The time-dependent equations in
the model are the Cahn-Hillliard equation for Li-diffusion:}{\small \par}

\noindent \textcolor{black}{\small{}
\begin{align}
\frac{\partial c}{\partial\tau} & =\nabla^{2}\frac{\delta\mathcal{F}}{\delta c},\label{eq:5}
\end{align}
}{\small \par}

\textcolor{black}{\small{}and the elastic-relaxation equation for
the host lattice structure:}{\small \par}

\noindent \textcolor{black}{\small{}
\begin{align}
\frac{\partial\psi}{\partial n} & =-\frac{\delta\mathcal{F}}{\delta\psi}+\frac{1}{L^{2}}\int\frac{\delta\mathcal{F}}{\delta\psi}d\vec{x}.\label{eq:6}
\end{align}
}{\small \par}

\noindent \textcolor{black}{\small{}In Eqs. \ref{eq:5} \textendash{}
\ref{eq:6}, we assume that the elastic relaxation (equilibrating
the peak density field) is infinitely faster than the evolution of
the composition field. Consequently, we model the equilibrium lattice
arrangements by maintaining $\frac{\delta\mathcal{F}}{\delta\psi}\approx0$
throughout the phase transition. The CH-PFC model is solved using
an Euler discretization scheme in a 2D finite-difference framework.
The computational grids have spacings of $\delta x=\delta y=\frac{4\pi}{q_{0}6\sqrt{3}}$
and have periodic boundary conditions.}\footnote{\noindent \textcolor{black}{\footnotesize{}The length scale of the
CH-PFC model is $\frac{1}{q_{0}}$.}}\textcolor{black}{\small{} At each grid point the values of the composition
field $c(\vec{x})$, the peak density field $\psi(\vec{x})$, and
the constant $r(\vec{x})$ are represented in discrete form, as $c_{ij}$,
$\psi_{ij}$ and $r_{ij}$ respectively. The time-derivative in Eq.
\ref{eq:5} is iterated at regular time intervals of $\Delta\tau=1$.
At each time step, the composition field $c_{ij}(\tau)$ is used to
update the lattice transformation coefficients, $\alpha(c_{ij}(\tau))$
and $\beta(c_{ij}(\tau))$. The coupled Laplacian $\nabla_{c}^{2}$
is next updated and the equilibrium lattice arrangements $\psi_{ij}$
are computed following Eq. \ref{eq:6}. This general numerical procedure
produces microstructural evolution and phase transitions.}{\small \par}
\begin{singlespace}

\section*{{\small{}Application to electrode microstructures}}
\end{singlespace}

\textcolor{black}{\small{}In this section, we model a representative
electrode-reservoir system. The system comprises a polycrystalline
FePO$_{4}$ electrode particle surrounded by an amorphous Li-reservoir.
We compute an electrochemical cycle by inserting/extracting Li-ions
into interstitial sites of the FePO$_{4}$ lattice structure. Here,
we study how lattices distort across a diffuse phase-boundary, and
investigate grain-boundary migration.}{\small \par}

\textcolor{black}{\small{}}
\begin{figure}
\begin{centering}
\textcolor{black}{\small{}\includegraphics[width=1\columnwidth]{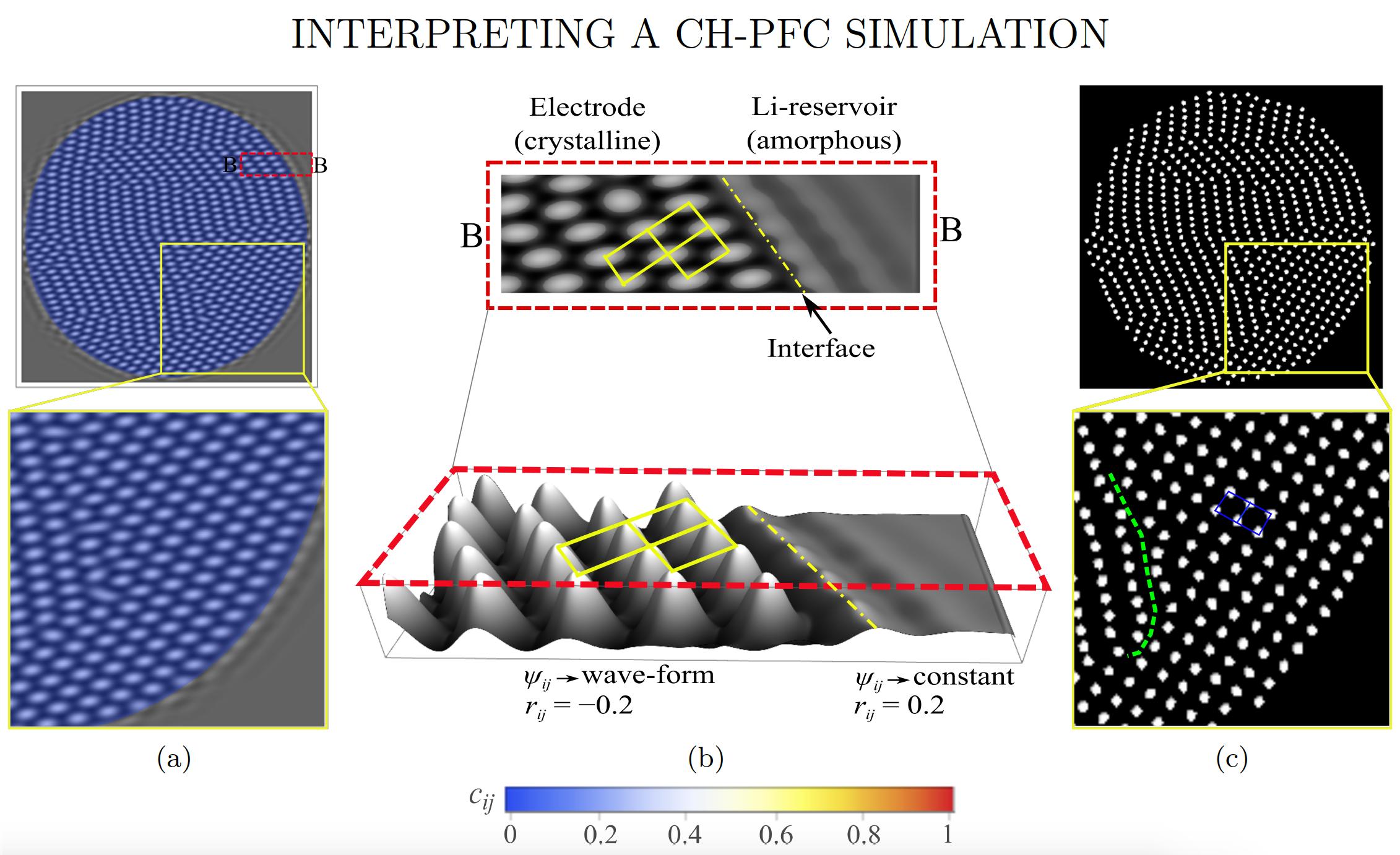}}
\par\end{centering}{\small \par}
\textcolor{black}{\small{}\caption{{\footnotesize{}\label{fig:2}(a) A CH-PFC simulation of a polycrystalline
FePO$_{4}$ electrode particle surrounded by an amorphous-Li-reservoir.
The inset box (below) shows the host lattice points (ellipsoidal peaks)
describing the FePO$_{4}$ lattice motif. The color bar indicates
Li-composition $c_{ij}$ in the electrode's interstitial sites. (b)
An enlarged image of the electrode lattice symmetry is shown in the
inset box BB. In the electrode region, the peak density field $\psi_{ij}$
has a wave form where the coordination symmetry of ellipsoidal peaks
represent an FePO$_{4}$ lattice. In the Li-reservoir region, $\psi_{ij}$
is a constant and describes an amorphous state. At the electrode-reservoir
interface, $\psi_{ij}$ changes as a function of $r_{ij}$ according
to Eq. }\textcolor{black}{\footnotesize{}\ref{eq:1}}{\footnotesize{}.
(c) ``Peak marker image'' illustrating the centroids of the ellipsoidal
peaks described by $\psi_{ij}$ (in Fig. }\textcolor{black}{\footnotesize{}\ref{fig:2}a)}{\footnotesize{}.
The inset box highlights the lattice structure (solid-blue lines)
and grain boundaries (dashed-gre}\textcolor{black}{\footnotesize{}en
lines) in the FePO$_{4}$ electrode. Details on the crystallographic
features of the polycrystalline electrode are shown in Fig. \ref{fig:7}. }}
}{\small \par}
\end{figure}
{\small \par}
\begin{singlespace}

\subsection*{{\small{}Electrode-reservoir system}}
\end{singlespace}

\textcolor{black}{\small{}A circular electrode of diameter $d=290\delta x$
is modeled on a periodic grid of size $300\delta x\times300\delta y$.
The electrode is surrounded by a reservoir-region, see Fig. \ref{fig:2}a.
On the computation grid $r_{ij}$ is locally defined to model a crystalline-region
for the electrode $(r_{ij}=-0.2)$ and an amorphous-region for the
reservoir $(r_{ij}=0.2)$. A homogeneous composition $c_{ij}=0$ is
described on the computation grid. The transformation coefficients
define the coarse-grained bond lengths and bond angles of the FePO$_{4}$
lattice geometry (see Table \ref{tab:1}) as inputs to the CH-PFC
simulation. At the initial state of the computation, a random peak
density field value, $-0.1\leq\psi_{ij}\leq0.5$, is assigned at each
grid point. Using Eq. \ref{eq:6}, the peak density field for the
electrode-reservoir system is computed. The peak density field, in
the electrode region, nucleates into FePO$_{4}$ crystallites with
different orientations. The initial condition with $-0.1\leq\psi_{ij}\leq0.5$
allows grains to evolve independently and form a polycrystalline electrode
as shown in Fig. \ref{fig:2}a. }{\small \par}

\textcolor{black}{\small{}In Fig. \ref{fig:2}(a-b), the peak density
field $\psi_{ij}$ has a wave-form in the electrode region and represents
the FePO$_{4}$ lattice motif, see inset-box. The composition field
for Li interstitials is $c_{ij}=0$. In the reservoir region, $\psi_{ij}$
is a constant and represents an amorphous state. At the electrode-reservoir
interface, fading bands of the peak density field are observed, see
inset box in Fig. \ref{fig:2}b. These bands show a gradual change
in $\psi_{ij}$, from its periodic wave-form in the crystalline electrode,
to a constant value in the amorphous reservoir. This change in $\psi_{ij}$
results from the abrupt change of $r_{ij}$ values modeled across
the electrode-reservoir interface. In this paper, the fading bands
of $\psi_{ij}$ represent the amorphous phase gradually taking on
a crystalline character. }{\small \par}

\textcolor{black}{\small{}In Fig. \ref{fig:2}a the peaks are of ellipsoidal
shape. This shape results from the coordinate transformation coefficients
in the Laplace operator $\nabla_{c}^{2}$, which shears the peaks
anisotropically. Fig. \ref{fig:2}a shows multiple grain boundaries
in the FePO$_{4}$ electrode particle. The peaks at the grain boundaries,
appear smeared and deviate from the conventional ellipsoidal shape.
We interpret this smearing of peaks as a coarse-grained lattice-structure
distortion, which maintains coherency between neighboring grains. }{\small \par}

\begin{figure}[h]
\begin{centering}
\textcolor{black}{\small{}\includegraphics[width=0.5\columnwidth]{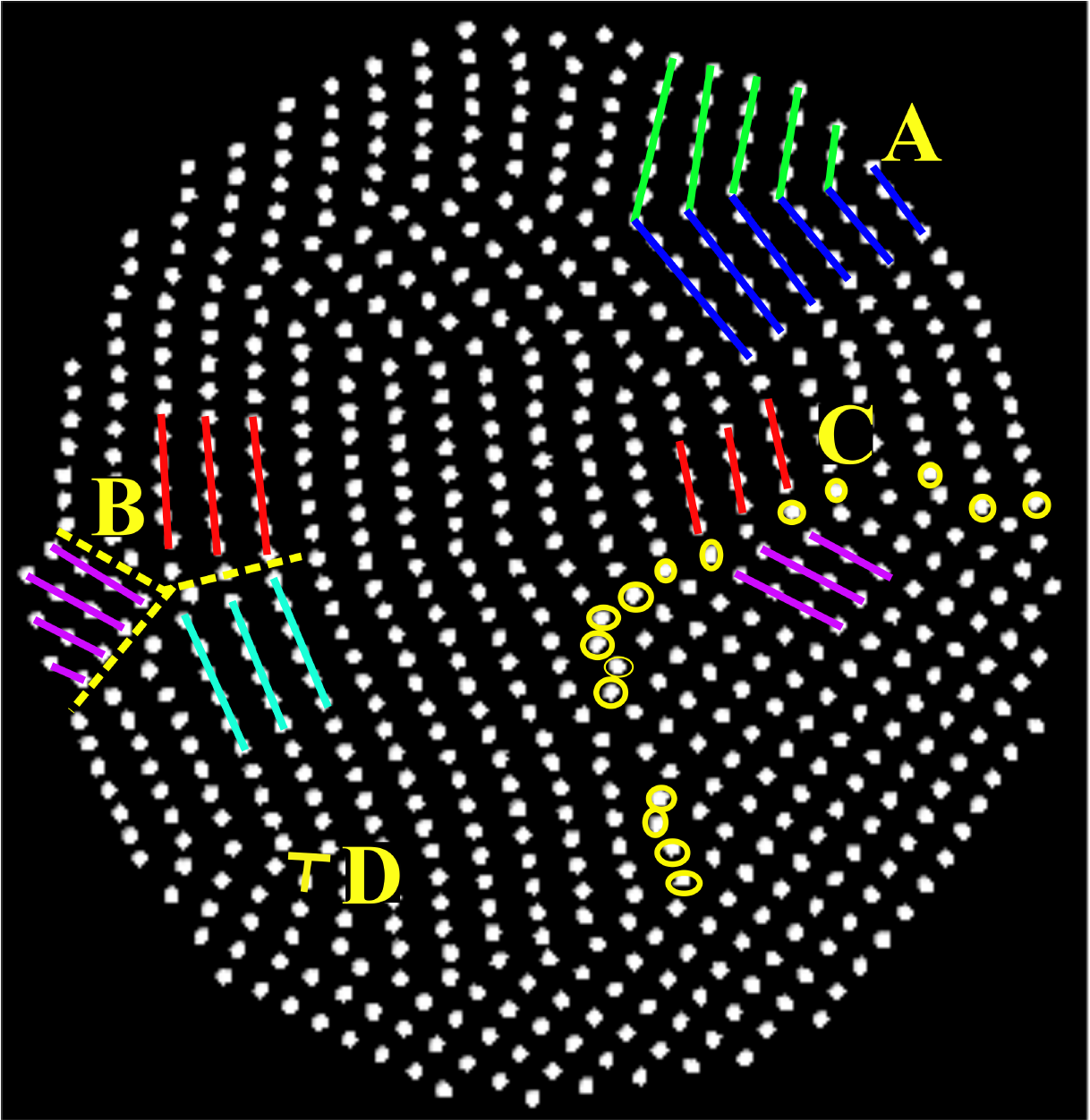}}
\par\end{centering}{\small \par}
\textcolor{black}{\small{}\caption{{\footnotesize{}\label{fig:7}An illustration of the }\textcolor{black}{\footnotesize{}crystallographic
features in the polycrystralline FePO$_{4}$ electrode particle (shown
in Fig. \ref{fig:2}a). (A) A coherent grain boundary with a perfect
match between the crystallographic planes. (B) A triple junction formed
at the intersection of three grains. The dashed yellow lines highlight
the grain boundary }{\footnotesize{}system. (C) A chain of defects,
highlighted by yellow circles, form a high angle grain boundary $(>15^{\circ})$.
(D) An edge-dislocation defect illustrated by a `T'. }}
}{\small \par}
\end{figure}

\textcolor{black}{\small{}The centroids of the ellipsoidal peaks describe
the rectangular coordination of the FePO$_{4}$ lattice. To highlight
this symmetry, we substitute the ellipsoidal peaks with a Gaussian
distribution \textendash{} we refer to this image as a ``peak marker
image'', see Fig. \ref{fig:2}c. Here, the centroids of the peaks,
$(x_{0},y_{0})$ are identified, and a 2D Gaussian distribution, $g(x,y)=\frac{2}{\pi}\exp[-2\{(x-x_{0})^{2}-(y-y_{0})^{2}\}]$
is modeled around each centroid. In Fig. \ref{fig:2}c, these Gaussian
distributions are shown by white solid dots. The fading bands of $\psi_{ij}$
at the electrode-reservoir interface are removed in the peak marker
image. The FePO$_{4}$ lattice structure and a representative grain
boundary are traced by solid blue and dashed green lines respectively,
see the inset box in Fig. \ref{fig:2}c. Fig. \ref{fig:7} highlights
the crystallographic features in the FePO$_{4}$ electrode particle.
Note, the peak-positions in Fig. \ref{fig:2}c does not correspond
to the atomic sites. The arrangement of peaks illustrates the coarse-grained
lattice-symmetry of the electrode particle. The grain boundaries arise
from the coarse-grained lattice-distortions in the underlying atomic
arrangements. Further details on the interpretation of the coarse-grained
lattices in CH-PFC simulations is provided in the appendix of Ref.
\cite{Balakrishna_Carter_2017}.}{\small \par}

\noindent \textcolor{black}{\small{}}
{\small \par}

\textcolor{black}{\small{}In Fig. \ref{fig:7}, we identify four representative
crystallographic features `A-D' of the FePO$_{4}$ electrode particle
(simulated in Fig. \ref{fig:2}a). These crystallographic features
arise naturally from computing the time derivative of $\psi_{ij}$
in Eq. \ref{eq:6}. Label `A', highlighted by the green and blue crystallographic
planes, marks a coherent interface. These coherent grain boundaries
are commonly observed in our CH-PFC simulations. Label `B' marks three
grain boundaries that intersect to form a triple junction. The dihedral
angle(s) between these grain boundaries influence(s) the migration
of the grain boundary system. For example, if the grain boundaries
are equally inclined at $120^{\circ}$, the triple-junction tends
not to move. However, if the grain boundaries are unequally inclined,
they drag the triple-junction towards the centre of a single grain.
This ``triple-junction drag-effect'' is discussed in the next section
(Electrochemical cycle). Labels `C' and `D' indicate, respectively,
the high-angle and low-angle grain boundaries in the electrode particle.
Grain boundaries with larger misorientation angle $>15^{\circ}$ are
incoherent and have a chain of edge-dislocations as shown by label
`C'. Grain boundaries with a misorientation angle less than $15^{\circ}$
tend to be more coherent and have fewer defects. For example, in Fig.
\ref{fig:7}, `D' shows an edge-dislocation defect that is highlighted
by a pair of short yellow lines. We next lithiate and delithiate this
polycrystalline electrode particle to investigate how its crystallographic
features evolve during an electrochemical cycle.}{\small \par}
\begin{singlespace}

\subsection*{{\small{}Electrochemical cycle }}
\end{singlespace}

\noindent \textcolor{black}{\small{}In the following simulations,
we specify a chemical potential to the Li-reservoir, which is held
fixed during the lithiation / delithiation cycle. This boundary condition
is a proxy for our assumption that the Li-diffusion rate is greater
in the reservoir than in the electrode particle. For example, we specify
the Li-reservoir with $c_{ij}=1$ that is fixed during lithiation.
The composition gradient between the FePO$_{4}$ electrode $(c_{ij}=0)$
and the Li-reservoir creates a boundary condition, which causes Li-ions
to diffuse into the electrode; see Fig. \ref{fig:3}. }{\small \par}
\noindent \begin{center}
\textcolor{black}{\small{}}
\begin{figure}
\begin{centering}
\textcolor{black}{\small{}\includegraphics[width=0.8\columnwidth]{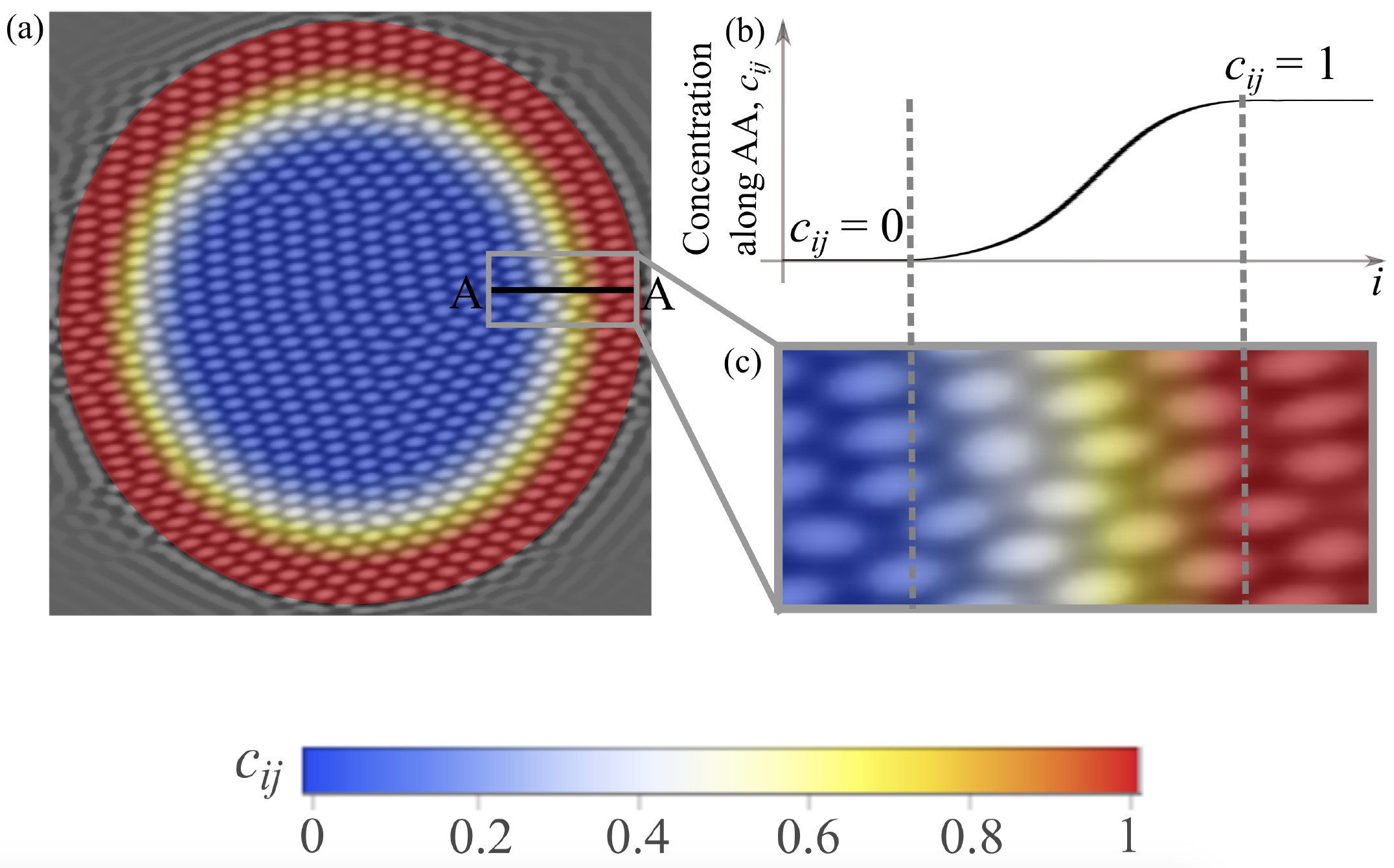}}
\par\end{centering}{\small \par}
\textcolor{black}{\small{}\caption{{\footnotesize{}\label{fig:3}(a) Lattice arrangements in a partially
lithiated FePO$_{4}$ electrode particle. The color bar illustrates
Li-composition in FePO$_{4}$ (blue $c_{ij}=0$) and LiFePO$_{4}$
(red $c_{ij}=1$) phases. A diffuse phase boundary (yellow $0<c_{ij}<1$)
separates the two phases. (b) The distribution of Li-composition field
$c_{ij}$ across a diffuse phase boundary (for example, along AA).
(c) The peak density field $\psi_{ij}$ across the diffuse phase-boundary.
The width of the diffuse interface (highlighted by vertical dashed-lines)
is numerically calibrated to span over $\sim4$ lattice points. }}
}{\small \par}
\end{figure}
\par\end{center}{\small \par}

\textcolor{black}{\small{}Fig. \ref{fig:3}a is a CH-PFC simulation
of a partially lithiated electrode particle. Fig. \ref{fig:3}b shows
the composition variation across the diffuse FePO$_{4}$ / LiFePO$_{4}$
phase boundary. The inset image in Fig. \ref{fig:3}c illustrates
the coarse-grained lattice distortions across the phase boundary.
This phase boundary is analogous to the `staging structure' observed
in some intercalation compounds \cite{staging_FPLFP}. For example,
the interfacial region between the FePO$_{4}$ and LiFePO$_{4}$ phases
has non-stoichiometric Li-composition (i.e., Li$_{X}$FePO$_{4}$
where $0<X<1$). }{\small \par}
\noindent \begin{center}
\textcolor{black}{\small{}\medskip{}
}
\par\end{center}{\small \par}

\begin{singlespace}
\noindent Lithiation 
\end{singlespace}

\textcolor{black}{\small{}Fig. \ref{fig:4} shows the temporal evolution
of the electrode microstructure during lithiation, as a function of
$c_{ij}$ and $\psi_{ij}$. The subfigures in row 1 illustrate the
lattice arrangements in the host-electrode particle as a function
of the interstitial Li-composition. The corresponding peak marker
images of the host-electrode are shown in row 2. The approximate positions
of three representative grain boundaries `A-C' are marked on the electrode's
lattice structure. The positions and orientations of these grain boundaries
are tracked during the electrochemical cycle. Subfigures in row 3
are the distortion maps that indicate the absolute difference in electrode's
peak positions with a reference state. That is, the host-electrode
lattices undergo displacive transformations during an electrochemical
cycle, which are tracked on the distortion maps. In the lithiation
cycle, the reference state is the initial FePO$_{4}$ particle shown
in Fig. \ref{fig:4}a. The distortion at each grid point is calculated
as $\delta_{ij}=\frac{\left\Vert x_{ij}(\tau)-x_{ij}(\tau=0)\right\Vert }{\delta_{0}}$
. Here, $x_{ij}(\tau)$ is the discrete representation of the peak
position at time $\tau$. The $\left\Vert \mathbf{...}\right\Vert $
represents the euclidean distance between a peak position at time
$\tau$ with reference to its initial position at $\tau=0$ in Fig.
\ref{fig:4}a. The normalizing constant, $\delta_{0}=\beta_{\mathrm{FP}}=\frac{4\pi}{q_{0}\sqrt{3}}$,
is the equilibrium separation between two adjacent Gaussian-peaks.
The distortion maps illustrate the lattice deformations in the host-electrode
particle that are induced by Li-diffusion. }{\small \par}

\textcolor{black}{\small{}As the Li-intercalation wave propagates
into the electrode as in Fig. \ref{fig:4}a \textendash{} \ref{fig:4}c,
the shape and size of grains `A-C' change. In the grain-boundary system
`A', the smallest grain `a$_{1}$' near the electrode edge shrinks
until it disappears in Fig. \ref{fig:4}d. This relieves the inhomogeneous
strain field in the larger grain `A'. The larger grain `A' grows in
size during lithiation. The grain boundary `C' tends to move towards
its center of curvature. This is consistent with the motion by curvature
observed in material microstructures \cite{Book_microstructural_evolution,GB_mechanism,Nestler book}.
In Fig. \ref{fig:4}b row 3, lattice distortions $\delta_{ij}\thickapprox0.12$
are observed in the lithiated phase of the electrode particle. We
interpret that these distortions correspond with the ferroelastic
lattice strains that accompany FePO$_{4}$ / LiFePO$_{4}$ phase transition.
The lattice arrangements in the electrode core show negligible variation
in peak positions, $\delta_{ij}\approx0$. In Fig. \ref{fig:4}c row
3, relatively large distortions $\delta_{ij}>0.5$ are observed along
grain boundaries in the electrode. We interpret these distortions
$\delta_{ij}>0.5$ to correspond with the grain-boundary migration.
At the end of lithiation, a LiFePO$_{4}$ phase forms in the host
electrode, see Fig. \ref{fig:4}d. The distortion map in Fig. \ref{fig:4}d
row 3, indicates ferroelastic lattice strains throughout the electrode
particle. }{\small \par}
\noindent \begin{center}
\textcolor{black}{\small{}}
\begin{figure}
\begin{centering}
\includegraphics[width=1\columnwidth]{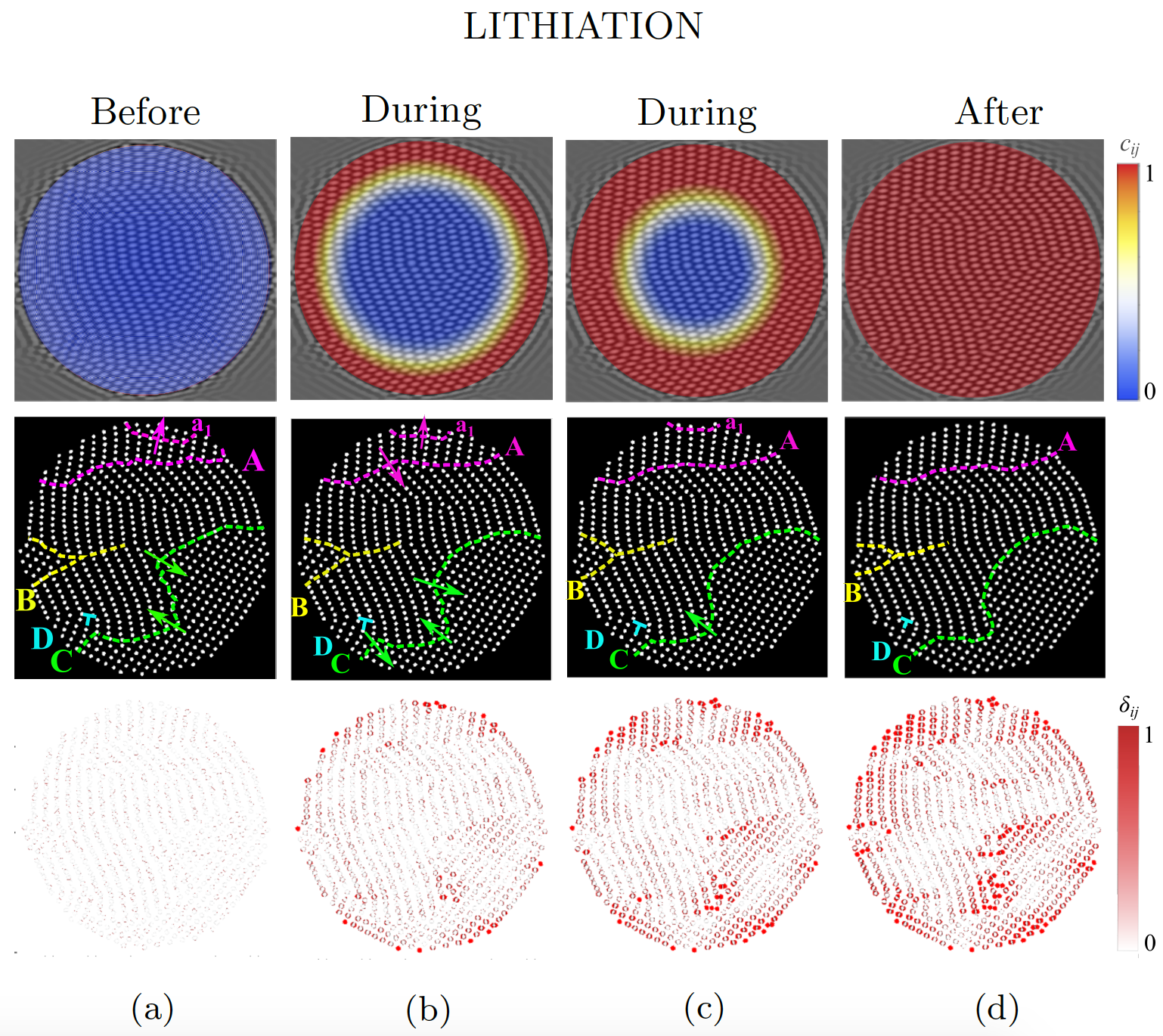}
\par\end{centering}
\caption{{\footnotesize{}\label{fig:4}(a-d): Lithiation of the polycrystalline
FePO$_{4}$ electrode particle. Starting from an initial FePO\protect\textsubscript{4}
phase (a), Li-ions intercalate into the electrode particle (b-c).
A LiFePO\protect\textsubscript{4} phase is formed at the end of lithiation
(d). The subfigures in row 1 show the temporal evolution of the Li-composition
$c_{ij}$ in the electrode particle. The subfigures on row 2 show
the structural transformation of lattices in the host-electrode during
lithiation. The labels `A\textendash D' highlight representative crystallographic
features in the electrode particle. The subfigures in row 3 illustrate
the distortion maps $\delta_{ij}$ corresponding to each stage of
lithiation. We interpret $\delta_{ij}$ to show host lattice distortions
induced from Li-intercalation.}}
\end{figure}
\par\end{center}{\small \par}

\begin{singlespace}
\noindent Delithiation
\end{singlespace}

\textcolor{black}{\small{}We next delithiate the electrode particle
starting from Fig. \ref{fig:4}d. We model the Li-reservoir with a
composition field $c_{ij}=0$, which is held fixed throughout the
delithiation process. Fig. \ref{fig:5} shows the temporal evolution
of the electrode microstructure as Li-ions are extracted from the
host lattices. The subfigures in row 1 show the microstructures as
a function of the interstitial Li-composition. The host-electrode
lattice arrangements and distortion maps are illustrated in row 2
and row 3, respectively. Note, the distortion maps in Fig. \ref{fig:5}
plot the absolute difference in peak positions, using the lattice
arrangements in Fig. \ref{fig:5}a as the reference. }{\small \par}
\begin{center}
\begin{figure}
\begin{centering}
\textcolor{black}{\small{}\includegraphics[width=1\columnwidth]{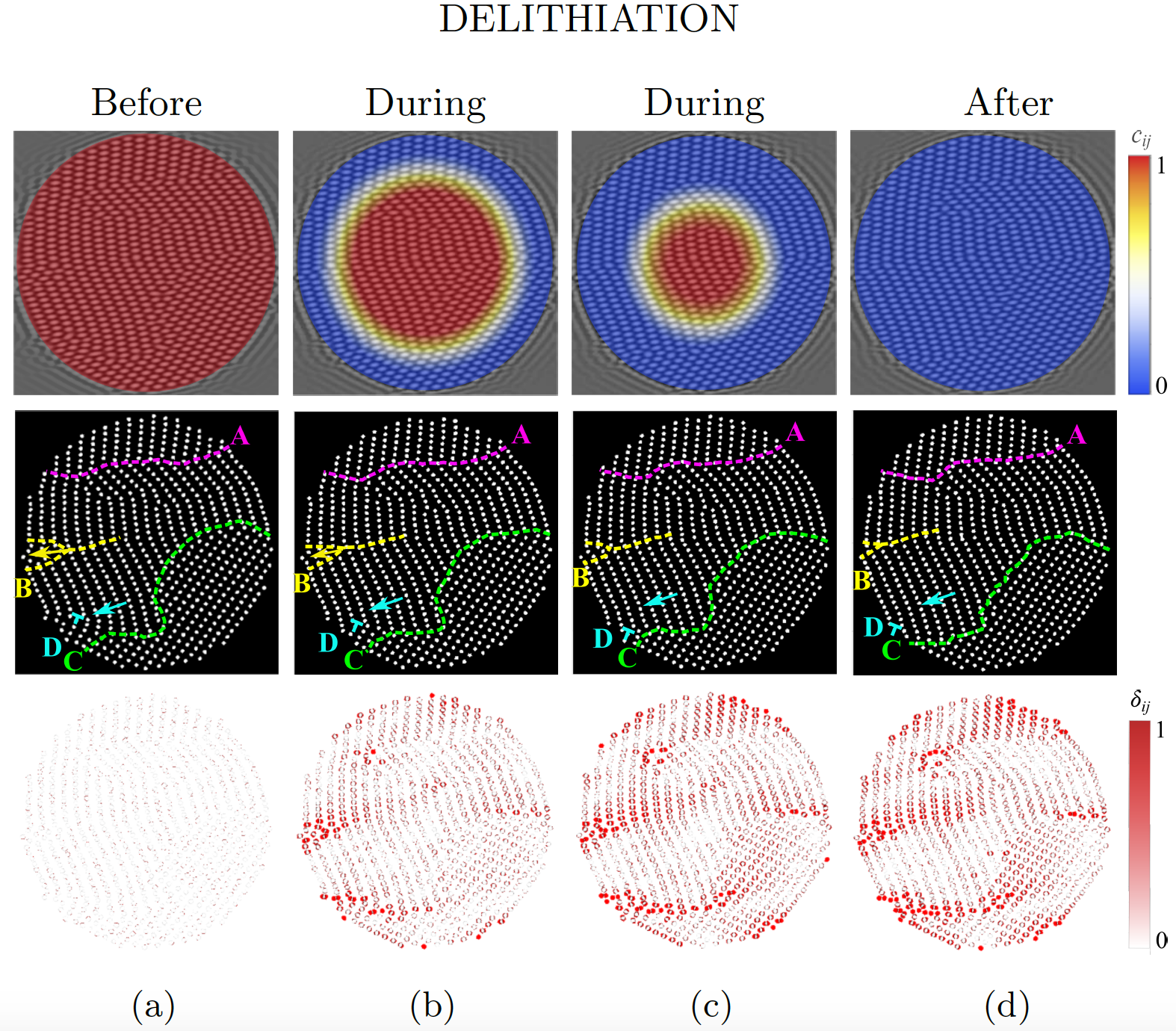}}
\par\end{centering}{\small \par}
\textcolor{black}{\small{}\caption{{\footnotesize{}\label{fig:5}(a-d):}\textbf{\footnotesize{} }{\footnotesize{}Delithiation
of a LiFePO$_{4}$ electrode particle. Starting from the initial LiFePO\protect\textsubscript{4}
phase (a), Li-ions are extracted from the electrode particle (b-c).
An FePO\protect\textsubscript{4} phase is formed at the end of delithiation
(d). The subfigures in row 1 show the temporal evolution of Li-composition
$c_{ij}$ in the electrode particle. The subfigures on row 2 show
the structural transformation of lattices in the host-electrode during
delithiation. The labels `A\textendash D' highlight representative
grain boundaries. The subfigures in row 3 illustrate the lattice distortion
maps $\delta_{ij}$ corresponding to each stage of delithiation. The
distortions in Fig. }\textcolor{black}{\footnotesize{}\ref{fig:5}
}{\footnotesize{}are calculated as the absolute difference}\textcolor{black}{\footnotesize{}
in peak positions using the lattice arrangements in subfigure }\textcolor{black}{\small{}\ref{fig:5}a}\textcolor{black}{\footnotesize{}
as reference}{\footnotesize{}. }}
}{\small \par}
\end{figure}
\par\end{center}

\textcolor{black}{\small{}During delithiation, LiFePO$_{4}$ lattices
undergo displacive transformation to describe FePO$_{4}$ lattice
structure. Grain `B' shrinks in size and disappears at the end of
delithiation (Fig. \ref{fig:5}, row 2). The grain-boundary system
`B', with unequal dihedral angles, drags the triple junction as shown
in Fig. \ref{fig:5}(a-d). Note that grains `a$_{1}$' and `B' are
of comparable sizes. However, Fig. \ref{fig:4} \textendash{} \ref{fig:5}
qualitatively illustrates that grain `a$_{1}$' shrinks faster than
grain `B'. We interpret that the slow migration of the grain-boundary
system `B' results from the triple-junction drag effect. This drag
effect has been observed by Shvindlerman et al. \cite{Shvindlerman_paper,Shvindlerman}
and has also been reported in electrode microstructures \cite{Balke_Kalinin}.
The edge dislocation `D' moves towards the electrode particle's surface
during delithiation. Extraction of Li-ions from the host-electrode
induces this dislocation movement. In Fig. \ref{fig:5}d, the electrode
particle is returned to the FePO$_{4}$ phase at the end of delithiation.
The electrode particle in Fig. \ref{fig:5}d has a different lattice-arrangement
from its initial state in Fig. \ref{fig:4}a. Similar dislocation
dynamics have been observed in experiments \cite{LiNiCoO5},\cite{Dionne_group}.
For example, Ulvestad et al. \cite{LiNiCoO5} reported that an applied
current induced dislocation movement in an intercalation cathode particle
(LiNi$_{0.5}$Mn$_{1.5}$O$_{4}$). The dislocation was stable at
room temperature and migrated to the particle surface under an electrical
load \cite{LiNiCoO5}. In hydrogen-palladium nanocubes \cite{Dionne_group},
Narayan et al. \cite{Dionne_group} reported that the diffusion of
hydrogen-atoms in host-palladium particle mobilized its edge-dislocations. }{\small \par}

\textcolor{black}{\small{}In Figs. \ref{fig:4} \textendash{} \ref{fig:5}
grain-boundary migration $(\delta_{ij}\approx1)$ is primarily observed
in electrode regions that are swept across by the Li-intercalation
wave. We hypothesize that the grain-boundary migration is induced
by Li-intercalation \cite{WCC_Handwerker,Handwerker_Cahn}. That is,
insertion or extraction of Li-ions from the host lattice is accompanied
by a change in the host-lattice geometry. The difference in lattice
parameters between the lithiated and delithiated phases generates
lattice misfit strains. This lattice misfit results in a stored elastic
energy, which acts as an additional driving force for grain-boundary
migration beyond its curvature \cite{WCC_Handwerker,Handwerker_Cahn}.
We next test this hypothesis by simulating grain growth in the same
electrode particle (Fig. \ref{fig:4}a), without Li-intercalation.
In this non-cycled electrode particle, grain growth has no additional
driving force.}{\small \par}
\begin{center}
\textcolor{black}{\small{}}
\begin{figure}
\begin{centering}
\textcolor{black}{\small{}\includegraphics[width=0.9\columnwidth]{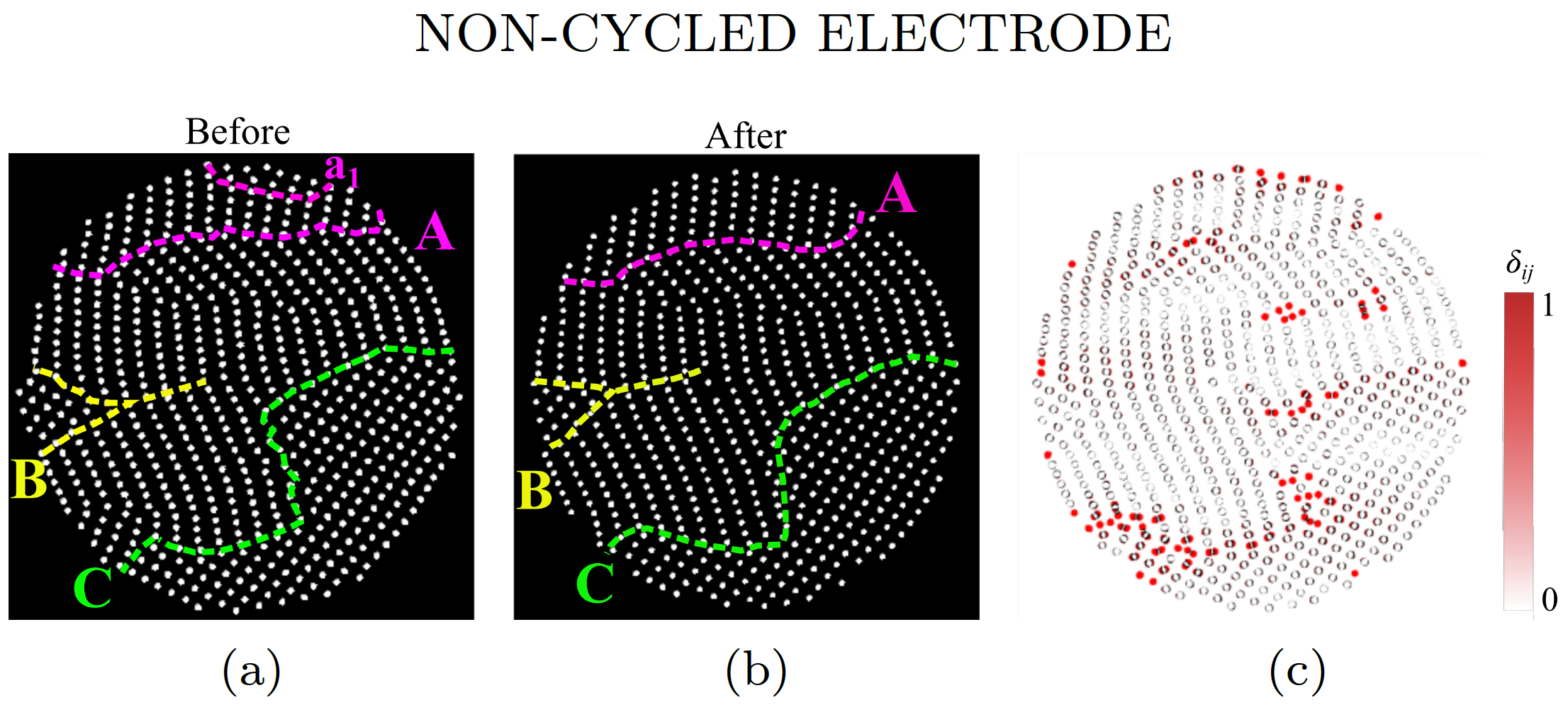}}
\par\end{centering}{\small \par}
\textcolor{black}{\small{}\caption{{\footnotesize{}\label{fig:6} Lattice rearrangements in the non-cycled
polycrystalline electrode particle. The electrode-reservoir system
is modeled with a homogeneous Li-composition $c_{ij}=0$, and computed
for the same time $(\tau_{\mathrm{total}})$ and temperature as in
Figs.}\textcolor{black}{\footnotesize{} \ref{fig:4} \textendash{}
\ref{fig:5}}{\footnotesize{}. The host-electrode lattice arrangements
(a) before $\tau/\tau_{\mathrm{total}}=0$, and (b) after $\tau/\tau_{\mathrm{total}}=1$
the non-cycled grain growth. (c) The lattice distortion map $\delta_{ij}$
calculated as the absolute difference in peak positions between the
`before' and `after' stages of the non-cycled electrode. The distortion
map illustrates lattice rearrangements near grain-boundaries. }}
}{\small \par}
\end{figure}
\par\end{center}{\small \par}

\begin{singlespace}
\noindent Non-cycled electrode
\end{singlespace}

\textcolor{black}{\small{}We model the electrode-reservoir system
as shown in Fig. \ref{fig:4}a. We define a homogeneous Li-composition
field $c_{ij}=0$ on the computation grid. We compute the time evolution
of lattice arrangements in the electrode following Eq. \ref{eq:6}
\cite{Balakrishna_Carter_2017}. We iterate Eq. \ref{eq:6} for the
same computation time $\tau_{\mathrm{total}}$ as required to electrochemically
cycle the electrode in Figs. \ref{fig:4} \textendash{} \ref{fig:5}.
We refer to this computation as the ``non-cycled grain growth''
of the electrode particle.}{\small \par}

\textcolor{black}{\small{}Fig. \ref{fig:6}a and Fig. \ref{fig:6}b
respectively illustrate the lattice arrangements in the host-electrode
particle before and after the computation. The lattice distortion
map in Fig. \ref{fig:6}c illustrates structural rearrangements in
the host electrode particle. Lattice distortions $\delta_{ij}\approx1$
are primarily observed along grain boundaries with a non-zero radius
of curvature. This suggests that grain-boundary migration in Fig.
\ref{fig:6}b \textendash{} \ref{fig:6}c is curvature-driven. These
migrations are however small in comparison to Figs. \ref{fig:4} \textendash{}
\ref{fig:5}. For example in Figs. \ref{fig:4} \textendash{} \ref{fig:5},
the grain `B' shrinks at the end of lithiation / delithiation cycles.
However in Fig. \ref{fig:6}, with no Li-intercalation, grain `B'
only slightly reduces in size. Similarly grains `A' and `C' show a
small variation in their shape and size. Fig. \ref{fig:6} is consistent
with our hypothesis that Li-intercalation assists (or accelerates)
grain-boundary migration.}{\small \par}
\begin{singlespace}

\section*{{\small{}Discussion}}
\end{singlespace}

\textcolor{black}{\small{}We compare grain growth in electrode particles
that were electrochemically cycled and non-cycled in Fig. \ref{fig:8a}(a-b),
respectively. The schematics, at the top of the plots show Li-composition
$c$ in the electrode particle at each computation time step, $\tau/\tau_{\mathrm{total}}$.
Note, $\tau_{\mathrm{total}}$ is the computation time required for
one electrochemical cycle. The grain size $\frac{\mathcal{A}}{\mathcal{A}_{e}}$
is normalized by the electrode area $\mathcal{A}_{e}$. Fig. \ref{fig:8a}a
illustrates that the mean grain-size, averaged over all the grains
in the electrode particle, grows by $11\%$ upon electrochemical cycling.
Smaller grains `a$_{1}$' and `B' shrink, and the larger grain `C'
has an approximately constant grain-size throughout the electrochemical
cycle. }{\small \par}

\textcolor{black}{\small{}In Fig. \ref{fig:8a}b, the mean grain-size
grows by $\sim2\%$ in the non-cycled electrode particle. The grain
sizes of A, B and C are approximately constant throughout the computation.
In the absence of Li-intercalation, the grain-boundary migration is
curvature-driven. In Fig. \ref{fig:8a}b, the grain growth (mean)
follows a relationship of the form $\frac{\mathcal{A}}{\mathcal{A}_{e}}\approx g_{0}+g(\frac{\tau}{\tau_{\mathrm{total}}})^{n}$,
where $n\approx0.5$.}\footnote{\noindent \textcolor{black}{\footnotesize{}$g_{0}=9.23,g=1.12$}}\textcolor{black}{\small{}
This is consistent with the curvature-driven grain growth observed
in experiments \cite{Book_microstructural_evolution}. In Fig. \ref{fig:8a}a
the grain growth follows a similar relationship, however, with the
exponent $n>0.5$. This suggests an accelerated grain growth (mean)
in Fig. \ref{fig:8a}a. Fig. \ref{fig:8a}(a-b) indicates that Li-intercalation
accelerates grain-boundary migration in the electrode particle. This
observation is consistent with the diffusion-induced grain-boundary
migration discussed in the works of Handwerker et al. \cite{Handwerker_Cahn,WCC_Handwerker}. }{\small \par}
\begin{center}
\textcolor{black}{\small{}}
\begin{figure}
\begin{centering}
\textcolor{black}{\small{}\includegraphics[width=1\columnwidth]{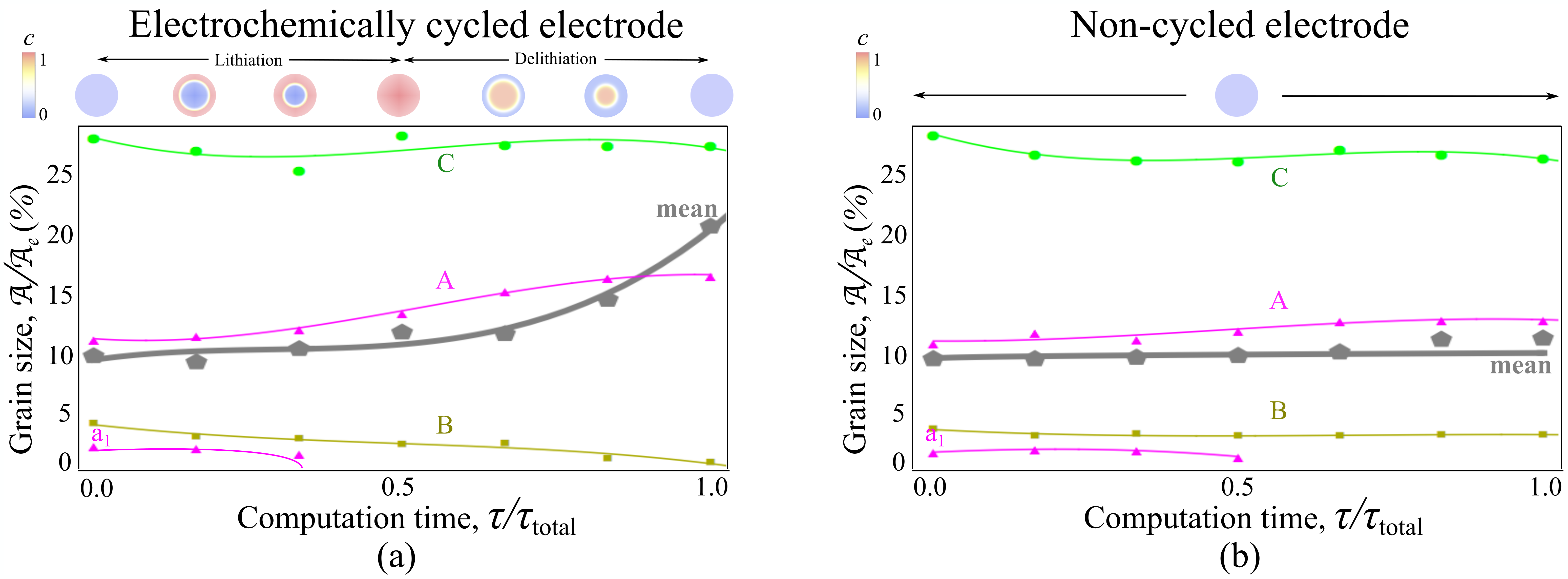}}
\par\end{centering}{\small \par}
\textcolor{black}{\small{}\caption{{\footnotesize{}\label{fig:8a}Grain growth $\mathcal{A}/\mathcal{A}_{e}$
in an electrode particl}\textcolor{black}{\footnotesize{}e when it
is (a) electrochemically cycled and (b) non-cycled. The schematics
above the plots illustrate Li-composition $c$ in the electrode particle
at the corresponding computation times. }}
}{\small \par}
\end{figure}
\par\end{center}{\small \par}

\textcolor{black}{\small{}We find the work of Bates et al. \cite{Bates_Dudney_2000}
on sputter-deposited thin film cathodes, as the closest experimental
comparison to the diffusion-induced grain growth effect suggested
by our present work. Bates et al. \cite{Bates_Dudney_2000} described
the cathodes as undergoing grain growth during electrochemical cycling
at near-ambient temperature \cite{Bates_Dudney_2000}. Another related
phenomena are the observations of accelerated grain growth in fluorite
structure solid electrolytes at elevated temperature. In these solid
electrolytes electrical load has been observed to accelerate grain
growth kinetics \cite{Yttria_GB_mobility,YSZ_GB_mobility}. We suggest
that systematic experimental investigation of battery electrodes after
extensive cycling will reveal lattice-strain-induced grain growth
as observed in this paper. }{\small \par}
\begin{center}
\textcolor{black}{\small{}}
\begin{figure}[h]
\begin{centering}
\textcolor{black}{\small{}\includegraphics[width=1\columnwidth]{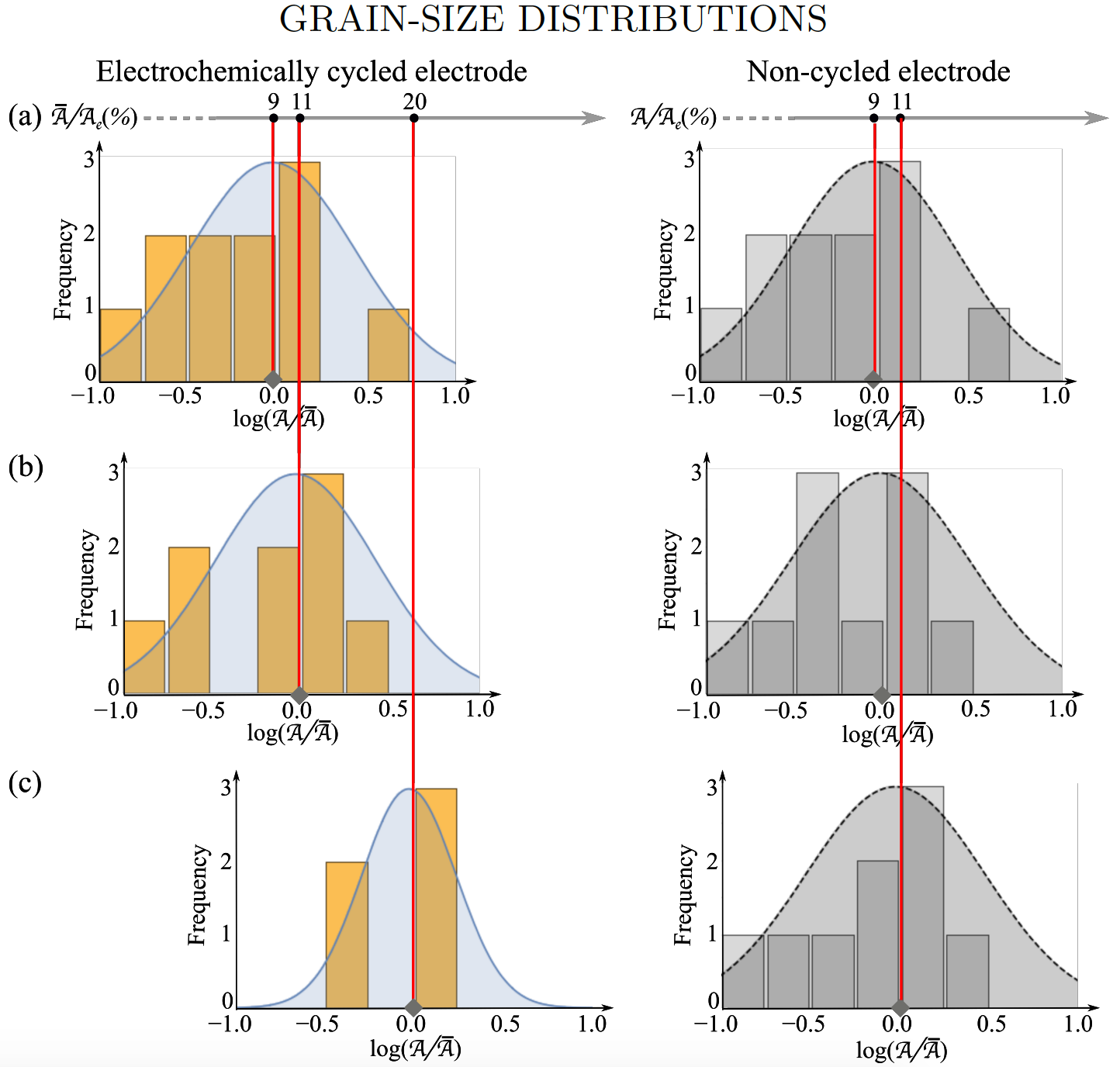}}
\par\end{centering}{\small \par}
\textcolor{black}{\small{}\caption{{\footnotesize{}\label{fig9}Grain-size distributions in the electrochemically
cycled (left) and non-cycled (right) electrode particle. The subfigures
correspond to the electrode microstructures at (a) $\tau/\tau_{\mathrm{total}}=0$
(b) $\tau/\tau_{\mathrm{total}}=0.5$ and (c) $\tau/\tau_{\mathrm{total}}=1.$
The position of the red lines corresponds to the mean grain-size values
on the vertical axes in Fig. }\textcolor{black}{\footnotesize{}\ref{fig:8a}.
}{\footnotesize{}The grain-size distributions indicate that }\textcolor{black}{\footnotesize{}electrochemical
cycling accelarates grain-growth in the electrode particle.}}
}{\small \par}
\end{figure}
\par\end{center}{\small \par}

\textcolor{black}{\small{}Fig. \ref{fig9}(a-c) shows the grain-size
distributions for the electrochemically cycled and non-cycled electrode
particle. Histograms of grain-size distributions at three stages of
the CH-PFC computation, namely $\tau/\tau_{\mathrm{total}}=0,0.5\ \mathrm{and\ }1$,
are plotted. They show the mean grain-size ($\bar{\mathcal{A}}/\mathcal{A}_{e}$)
and number of grains (frequency) in the electrode-particle at each
stage. The histograms in Fig. \ref{fig9}(a-c) show a good fit to
the log-normal size distributions, which are observed in experiments
\cite{Yttria_GB_mobility,YSZ_GB_mobility,Nestler}. In the electrochemically
cycled electrode, the variance (width) of the grain-size distribution
decreases upon Li-intercalation. In the non-cycled electrode particle,
see Fig. \ref{fig9}(right), the variance of its size-distribution
is approximately constant throughout the computation. These results
suggest that smaller grains in the electrochemically cycled electrode
particle, with $\log(\mathcal{A}/\bar{\mathcal{A}})<-0.5$, disappear
at the expense of bigger grains. Electrode particles with enhanced
grain sizes tend to be less tough in comparison to samples with smaller
grains \cite{Swallow,Woodford}. Fig. \ref{fig9} suggests that Li-intercalation
is an additional process that affects the mechanical reliability in
batteries.}{\small \par}

\textcolor{black}{\small{}We next discuss a few issues that arise
in interpreting the CH-PFC simulations in Figs. \ref{fig:4} to \ref{fig:6}.
For example, the electrode-particle volume (or area) is held constant
during the computation. However, the number density of peaks in the
electrode-particle is not conserved.}\footnote{\noindent \textcolor{black}{\footnotesize{}Please note, the addition/removal
of peaks does not affect the coarse-grained lattice structure described
by $\psi_{ij}$ \cite{Balakrishna_Carter_2017}. }}\textcolor{black}{\small{} That is, the peaks move in/out at the electrode-reservoir
interface to accommodate for the lattice-area change during phase
transformation. This ``artificial boundary condition'' applied to
the electode-reservoir interface does not impose volume-confinement
stresses on the electrode-particle. In future works, we propose to
introduce numerical correction terms \cite{Lowengrub-Voight} to Eq.
(6) to conserve the total number of peaks in the electrode region.}{\small \par}

\textcolor{black}{\small{}Another issue is related to the isotropic
Li-diffusion in the elecrode-particle. In olivine materials, such
as FePO$_{4}$, Li-ions preferentially diffuse through the host's
one-dimensional channels \cite{Ceder_anisotropic_Li}. In previous
continuum models \cite{13b  Tang_Carter-1,12 Carter_Tang,Huttin_Kamlah_2012,14 Bai_Bazant_2011},
researchers have modeled this preferential diffusion by using an anisotropic
diffusion coefficient. In these methods the crystallographic planes
of FePO$_{4}$ were typically oriented relative to a cartesian basis.
The anisotopic Li-diffusion was modeled with reference to the coordinate
axes. In the CH-PFC model, however, the crystallographic planes of
FePO$_{4}$ / LiFePO$_{4}$ are a natural outcome in the simulations.
We assume an isotropic bulk-diffusion coefficient $D$ in Eq. (2)
to model lithiation/delithiation; see Figs. \ref{fig:4} and \ref{fig:5}.
We do, however, see anisotropy in Li-diffusion arising from the coordinate-transformation-coefficients
in the coupled term $\nabla^{2}\gamma\frac{\psi}{2}\frac{\partial G(\nabla_{c}^{2})}{\partial c}\psi$.}\footnote{\textcolor{black}{\footnotesize{}Please see Eq. (8) in the appendix.}}\textcolor{black}{\small{}
This contribution is negligible as a result of the diffuse width ($\sim4$
coarse-grained lattices) of the phase boundary. The value of the constant
$\gamma$ also influences the anisotropy contribution from the coupled
term. In this paper we set $\gamma=\frac{\lambda^{2}q_{0}^{5}}{uF_{0}}=1$
for computational expediency. That is, the energy contribution from
the PFC equation is negligible (i.e., $\sim10^{-19}F_{0})$. Consequently,
the isotropic diffusion of the Li-intercalation wave is relatively
unaffected by the anisotropy in $\psi_{ij}$. However, if $\gamma$
is large the effects from material crystallography and lattice defects
are expected to appear in the simulations. }{\small \par}

\textcolor{black}{\small{}Overall, the CH-PFC methods has three specific
advantages in modeling phase transitions in intercalation electrodes:
First, the CH-PFC simulations provide qualitative insights on phenomenological
mechanisms across different length and time scales. For example, the
simulations describe Li-diffusion at a continuum scale and grain-boundary
migration at an atomistic scale. Second, the crystallographic planes
of FP/LFP lattices in the electrode particle are an emergent phenomenon
of the CH-PFC simulations \cite{Provatas_review}. The crystallographic
misorientations at grain boundaries follow naturally from the simulations.
Third, in the CH-PFC model the coarse-grained lattices distort independently
as a function of the interstitial Li-composition. That is, lattice
geometries are different in the lithiated and delithiated phases.
This difference generates coherency strains across the diffuse phase
boundary. The coherency strains in the CH-PFC simulations are a natural
outcome resulting from the different lattice geometries, and are not
modeled as an approximation from the composition field \cite{12 Carter_Tang,13b  Tang_Carter-1,15 Cogswell_Bazant_2012,Huttin_Kamlah_2012}.
This results in inhomogeneous strains across grain boundaries and
phase boundaries in a polycrystalline electrode. }\medskip{}

\noindent \textbf{\small{}Summary}{\small \par}

\smallskip{}

\noindent \textcolor{black}{\small{}We demonstrate that Li-diffusion
accelerates grain growth in intercalation compounds using FePO$_{4}$
/ LiFePO$_{4}$ as a model system. We present an application of the
Cahn-Hilliard \textendash{} phase-field-crystal methods to model phase
transitions in a polycrystalline FePO$_{4}$ electrode particle. This
modeling approach illustrates multiscale interactions between Li-diffusion
(continuum parameter) and structural transformations of host-lattices
(atomistic) during an electrochemical cycle. While grain growth from
cyclic intercalation has not been experimentally confirmed, our study
suggests that volume changes in electrode upon Li-intercalation accelerates
grain growth. The results qualitatively describe phenomenological
mechanisms in intercalation electrodes, such as edge-dislocation movement
and the triple-junction-drag effect. The CH-PFC model could be applied
to investigate chemo-mechanically coupled problems that involve solute-induced
phase transitions. }\medskip{}

\noindent \textbf{\small{}Acknowledgements}{\small \par}

\smallskip{}

\noindent \textcolor{black}{\small{}The authors gratefully acknowledge
the support by the grant DE-SC0002633 funded by the U.S. Department
of Energy, Office of Science, in carrying out this work. A. Renuka
Balakrishna also acknowledges the support of the Lindemann postdoctoral
fellowship.}\medskip{}

\noindent \textbf{\small{}Appendix}{\small \par}

\smallskip{}

\noindent \textbf{\small{}A.1 CH-PFC model details}{\small \par}

\smallskip{}

\noindent \textcolor{black}{\small{}In this section we provide a brief
explanation of the CH-PFC model. For the derivation of the CH-PFC
model please refer to Ref. \cite{Balakrishna_Carter_2017}. The CH-PFC
model describes the total free energy $F$, as a function of the composition
field $\overline{c}$ and the peak density field $\phi$:}{\small \par}

\noindent \textcolor{black}{\small{}
\begin{align}
F & =\int\{g(\overline{c})+\kappa|\nabla\overline{c}|^{2}+f(\phi)+\frac{\phi}{2}G(\nabla_{c}^{2})\phi\}d\vec{\mathbf{r}}\nonumber \\
 & =\int\{RT[\overline{c}\mathrm{ln}(\overline{c})+(1-\overline{c})\mathrm{ln}(1-\overline{c})]+\Omega\overline{c}(1-\overline{c})\label{eq:1-1}\\
 & ~~+\kappa|\nabla\overline{c}|^{2}+\frac{\phi}{2}(a\Delta T_{0}+\lambda(q_{0}^{2}+\nabla_{c}^{2})^{2})\phi+u\frac{\phi^{4}}{4}\}d\vec{\mathbf{r}}.\nonumber 
\end{align}
}{\small \par}

\noindent \textcolor{black}{\small{}Eq. \ref{eq:1-1} in its normalized
form is:}{\small \par}

\noindent \textcolor{black}{\small{}
\begin{align}
\mathcal{F}=\frac{F}{F_{0}} & =\int\{g(c)+|\nabla c|^{2}+f(\psi)+\frac{\psi}{2}G(\nabla_{c}^{2})\psi\}d\vec{x}\nonumber \\
 & =\int\{c\mathrm{ln}(c)+(1-c)\mathrm{ln}(1-c)+\Omega_{0}c(1-c)\label{eq:2-1}\\
 & ~~+|\nabla c|^{2}+\gamma\left(\frac{\psi}{2}(r+(1+\nabla_{c}^{2})^{2})\psi+\frac{\psi^{4}}{4}\right)\}d\vec{x},\nonumber 
\end{align}
}{\small \par}

\noindent \textcolor{black}{\small{}where $c=\frac{\overline{c}_{a}-\overline{c}}{\overline{c}_{a}-\overline{c}_{b}}$,
and $\psi=\phi\sqrt{\frac{u}{\lambda q_{0}^{4}}}$. The gradient energy
coefficient $\kappa=\frac{F_{0}}{(\overline{c}_{a}-\overline{c}_{b})^{2}}\left(\frac{16\pi\xi}{q_{0}\sqrt{3}}\right)^{2}$,
is numerically calibrated such that the width of the diffuse composition
interface spans over $\sim4$ lattice spacings described by the peak
density field, $\psi$. Note, $\frac{1}{q_{0}}$ is the length scale
of the CH-PFC model and $\xi$ is the scale factor that coarse-grains
the lattice units. The form of Eq. \ref{eq:1-1}-\ref{eq:2-1} is
similar to that in the work of Renuka-Balakrishna and Carter \cite{Balakrishna_Carter_2017}
where detailed explanations of the specific terms, constants $(\overline{c}_{a},\overline{c}_{b},\lambda,q_{0},u,\xi)$
and normalizations are provided. Note, the coefficients in the regular
solution model are the same as in Ref. \cite{13b  Tang_Carter-1}.
In Eq. \ref{eq:2-1}, $g(c)$ and $f(\psi)$ describe the homogeneous
energy contributions from the Cahn-Hilliard and phase-field-crystal
equations respectively. The composition gradient-energy is given by
$|\nabla c|^{2}$. The parameter $r$, controls the second-order phase
transition of the PFC model. The constant $\gamma=\frac{\lambda^{2}q_{0}^{5}}{uF_{0}}$
relates the free energy normalizations of the Cahn-Hilliard and the
PFC model. In this paper we set $\gamma=1$ and $\xi=1$ for computational
expediency. }{\small \par}

\noindent \textcolor{black}{\small{}The Cahn-Hilliard and the phase-field-crystal
models are coupled via the operator $G(\nabla_{c}^{2})=(1+\nabla_{c}^{2})^{2}$.
The composition field is coupled to the lattice symmetry via the Laplace
operator:}{\small \par}

\noindent \textcolor{black}{\small{}
\begin{equation}
\nabla_{c}^{2}=\xi^{2}((\mathrm{A}_{11}^{2}+\mathrm{A}_{12}^{2})\frac{\partial}{\partial x^{2}}+\mathrm{A}_{22}^{2}\frac{\partial}{\partial y^{2}}+2\mathrm{A}_{12}\mathrm{A}_{22}\frac{\partial}{\partial x\partial y}).\label{eq:3-1}
\end{equation}
}{\small \par}

\noindent \textcolor{black}{\small{}Here, $\mathrm{A_{\mathit{kl}}}$
are the elements of the transformation matrix and are described as
functions of the composition field:}{\small \par}

\noindent \textcolor{black}{\small{}
\begin{equation}
\mathbf{A\mathrm{(\mathit{c})}}=\left[{\begin{array}{cc}
\alpha(c) & \frac{-\alpha(c)}{\sqrt{3}}\\
0 & \frac{2\beta(c)}{\sqrt{3}}
\end{array}}\right].\label{eq:4-1}
\end{equation}
}{\small \par}

\begin{figure}
\begin{centering}
\textcolor{black}{\small{}\includegraphics[width=0.5\columnwidth]{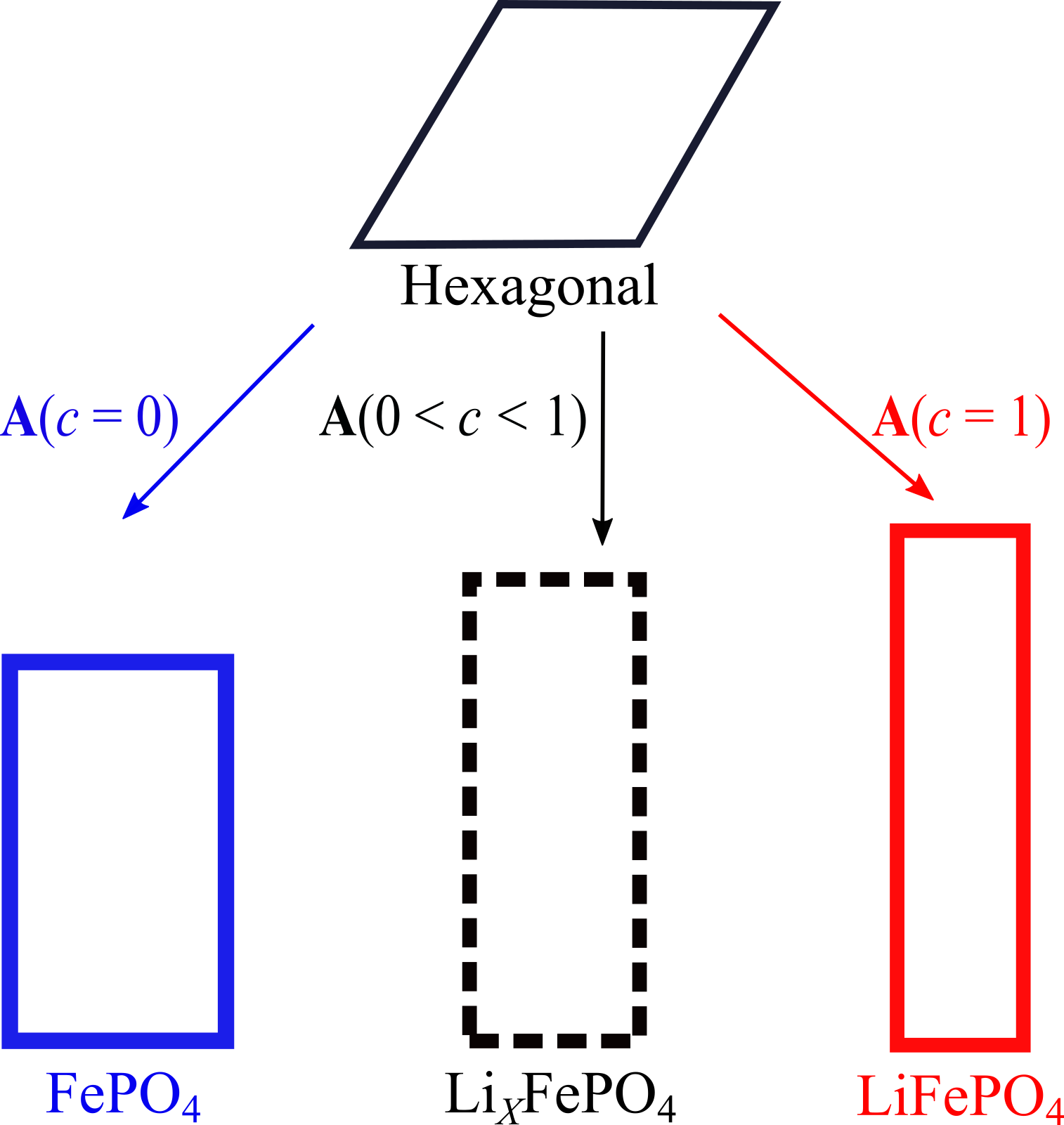}}
\par\end{centering}{\small \par}
\centering{}\textcolor{black}{\small{}\caption{{\footnotesize{}\label{fig:1}Schematic representations of the coupling
between lattice symmetry and composition field in the CH\textendash PFC
model. The transformation matrix $\mathbf{A}(c),$ is described as
a function of the composition field, $c$. For $c=0$, $\mathbf{A}(c)$
deforms a hexagonal symmetry to rectanglular FePO$_{4}$ lattice (in
blue, far left). For $c=1$, $\mathbf{A}(c)$ describes the lattice
motif of LiFePO$_{4}$ (in red, far-right). For $0<c<1$, intermediate
lattice geometries of the interphase Li$_{X}$FePO$_{4}$ (dashed
box) is modeled. }}
}{\small \par}
\end{figure}

\noindent \textcolor{black}{\small{}The matrix $\mathbf{A(\mathit{c})}$,
describes affine lattice transformations using hexagonal symmetry
as the reference structure. The transformation coefficients in Eq.
\ref{eq:4-1} transform a hexagonal symmetry (of unit size) to rectangular
geometries of FePO$_{4}$ (with $c=0)$ and LiFePO$_{4}$ (with $c=1$)
in 2D \cite{Padhi_Goodenough_1997}, see Fig. \ref{fig:1}. The transformation
coefficients are given by $\alpha(c)=\alpha_{\mathrm{FP}}+(\alpha_{\mathrm{LFP}}-\alpha_{\mathrm{FP}})c$
and $\beta(c)=\beta_{\mathrm{FP}}+(\beta_{\mathrm{LFP}}-\beta_{\mathrm{FP}})c$.
The values of $(\alpha_{\mathrm{FP}},\beta_{\mathrm{FP}})$ and $(\alpha_{\mathrm{LFP}},\beta_{\mathrm{LFP}})$
correspond to FePO$_{4}$ and LiFePO$_{4}$ lattices and are obtained
from Table \ref{tab:1}. During electrochemical cycling, the composition
field in a two-phase FP/LFP microstructure is a function $c(\vec{x})$.
The lattices in this microstructure are transformed according to $\mathbf{A}(c(\vec{x}))$. }{\small \par}

\noindent \textcolor{black}{\small{}}
{\small \par}

\textcolor{black}{\small{}Next, we list the detailed equations of
motion used in the CH-PFC model. The evolution of the composition
field follows a generalized Cahn-Hilliard equation:
\begin{align}
\frac{\partial c}{\partial\tau} & =\nabla^{2}\frac{\delta\mathcal{F}}{\delta c}\nonumber \\
 & =\nabla^{2}(\frac{\partial g(c)}{\partial c}-\nabla^{2}c+\gamma\frac{\psi}{2}\frac{\partial(\nabla_{c}^{4}+2\nabla_{c}^{2})}{\partial c}\psi).
\end{align}
}{\small \par}

\textcolor{black}{\small{}where, $\gamma=1$ and $\tau$ is the dimensionless
time variable $\tau=t\frac{D}{L^{2}}$. Here, $D$ is the isotropic
diffusion coefficient and $L$ is the size of the computational grid.
Eq. (8) introduces two Laplace operators, $\nabla^{2}$ and $\nabla_{c}^{2}$,
respectively. The Laplacian $\nabla^{2}=\frac{\partial^{2}}{\partial x^{2}}+\frac{\partial^{2}}{\partial y^{2}}$
describes an isotropic Li-diffusion. We do, however, see anisotopy
in $\frac{\partial c}{\partial\tau}$ because of the coordinate transformation
coefficients in $\nabla_{c}^{2}$ and from grain boundaries in the
host-electrode system. The coupled Laplacian $\nabla_{c}^{2}$ computes
the derivatives of the peak density field in a transformed coordinate
space. The term $\nabla^{2}\gamma\frac{\psi}{2}\frac{\partial(\nabla_{c}^{4}+2\nabla_{c}^{2})}{\partial c}\psi$
is anisotropic and influences the}\textbf{\textcolor{black}{\small{}
}}\textcolor{black}{\small{}Li-intercalation. This anisotropic effect
is negligible in the CH-PFC simulations of this paper, because the
diffuse phase boundary spans over $\sim4$ lattice spacings. The lattice
arrangement at each composition evolution step is computed as:}{\small \par}

\textcolor{black}{\small{}
\begin{align}
\frac{\partial\psi}{\partial n} & =-\frac{\delta\mathcal{F}}{\delta\psi}+\frac{1}{L^{2}}\int\frac{\delta\mathcal{F}}{\delta\psi}d\vec{x}\nonumber \\
 & =-\gamma[(r+(1+\nabla_{c}^{2})^{2})\psi+\psi^{3}]\\
 & ~~+\frac{1}{L^{2}}\int\gamma[(r+(1+\nabla_{c}^{2})^{2})\psi+\psi^{3}]d\vec{x}.\nonumber 
\end{align}
}{\small \par}

\textcolor{black}{\small{}Here, we assume that the elastic relaxation
(equilibrating the peak density field) is infinitely faster than the
evolution of the composition field. Consequently, we model the equilibrium
lattice arrangements by maintaining $\frac{\delta\mathcal{F}}{\delta\psi}\approx0$
throughout the phase transition. In Eq. (9), $n$ is a fictive time-like
variable that is rapidly changing in comparison to the dimensionless
time, $\tau$. Further details on Eq. (4) \textendash{} (5) can be
found in Ref. \cite{Balakrishna_Carter_2017}.}{\small \par}

\noindent \smallskip{}

\noindent \textbf{\small{}A.2 Modeling lattice symmetries using PFC
methods}{\small \par}

\smallskip{}

\noindent \textcolor{black}{\small{}The coordinate transformation
coefficients introduced in the CH-PFC model \cite{Balakrishna_Carter_2017}
resemble the lattice stretch and shear factors used in the anisotropic
PFC methods \cite{Kundin_Choudhary},\cite{APFC}. The stretch/shear
factors deform a lattice symmetry and are used to calculate the anisotropic
model coefficients. The transformation coefficients in the CH-PFC
model, however, differ from the anisotropic coefficients in two ways:
First, the CH-PFC model computes the Laplacian in a transformed coordinate
space. The coefficients correspond to the elements of a transformation
matrix, which relates lattice symmetries in 2D point groups via affine
transformations. Second, the transformation coefficients are coupled
to a composition field, which influences the underlying lattice symmetry
of the host material.}{\small \par}

\end{document}